\documentclass[10pt,a4paper,twoside]{article}
\usepackage{epsfig}
\usepackage{baltlat6}
\usepackage{array}
\usepackage{here}
\begin{document}
\ \
\vspace{0.5mm}
\setcounter{page}{1}
\vspace{8mm}

\titlehead{Baltic Astronomy, vol.\, , --, 2011}

\titleb{OBSERVATONS OF NGC 3077 GALAXY IN NARROW BAND [SII] AND H$\alpha$ FILTERS}

\begin{authorl}
\authorb{M. Andjeli\' c}{1} ,
\authorb{K. Stavrev}{2} ,
\authorb{B. Arbutina}{1} ,
\authorb{D. Ili\' c}{1} and
\authorb{D. Uro\v{s}evi\'{c}}{1}
\end{authorl}

\begin{addressl}
\addressb{1}{Department of Astronomy, Faculty of Mathematics,
University of Belgrade,\\ Studentski trg 16, Belgrade, Serbia;
 mandjelic@matf.bg.ac.rs}
\addressb{2}{Institute of Astronomy, Bulgarian Academy of Sciences,
 72 Tsarigradsko Shosse Blvd., BG-1784 Sofia, Bulgaria}
\end{addressl}

\submitb{Received: ; accepted: }

\begin{summary}

We present observations of the HI tidal arm near dwarf galaxy NGC 3077
(member of the M81 galaxy group) in narrow band [SII] and H$\alpha$ filters.
Observations were carried out in March 2011 with the 2m RCC telescope at NAO
Rozhen, Bulgaria. Our search for possible supernova remnant candidates (identified as
sources with enhanced [SII] emission relative to their H$\alpha$ emission) in this
region yielded no sources of this kind.
Nevertheless, we found a number of objects with significant H$\alpha$ emission
that probably represent uncatalogued, low brightness HII regions.

\end{summary}


\resthead{Observations of NGC 3077 galaxy in narrow band [SII] and H$\alpha$
 filters}
{M. Andjeli\' c, K. Stavrev, B. Arbutina, D. Ili\'c, \& D. Uro\v sevi\' c}

\sectionb{1}{INTRODUCTION}

The M81 galaxy group is the nearest interacting group of galaxies whose main members
are M81, M82 and NGC 3077. Yun et al. (1994) found prominent HI structures
surrounding these galaxies with large HI complexes and tidal bridges that are
a result of the galaxy encounters. VLA HI observatons (Walter et al. 2002) showed that
90\% of atomic hydrogen around NGC 3077 is located eastward of the galactic center,
in the tidal arm, called "Garland" (Barbieri et al. 1974, Karachentsev et al. 1985).
The estimated HI mass in NGC 3077 is about $10^8$ $\mathcal{M}_\odot$. Simulations
showed that the Garland was created by tidal disruptions between NGC 3077
and M81, about $3\times 10^8$ years ago (Thomasson \& Donner 1993, Yun 1999).

Tidal interaction between galaxies in this group is supposed to lead to
enhanced star formation. As a consequence, we expect to detect many supernova
remnant (SNR) and HII region candidates. Karachentsev et al. (2007)
reported that Garland has the highest star formation rate (SFR) per luminosity among 150 galaxies
of the Local Volume with known SFRs.

The aim of this work was to perform an optical search, using narrow [S II] and
H$\alpha$ filters, to identify SNR and HII region candidates in interaction regions
in the M81 galaxy group, particularly in the tidal feature Garland, near NGC 3077.
In this analysis we use the fact that
the optical spectra of SNRs have elevated [SII]/H$\alpha$ emission-line ratios
compared to the spectra of normal HII regions (Blair \& Long 2004,
Matonick \& Fesen 1997). This ratio is an accurate means of differentiating
between shock-heated SNRs (ratios $>$0.40, but often considerably higher)
and photoionized nebulae (0.40, but typically $<$0.2).

We observed a region of maximum HI emission located
in a prominent tidal arm, approximately 4 kpc east from the center of
the galaxy. So far, 36 HII regions have been detected in this region
(Walter et al. 2006), out of which 30 were in our observed field of view.
Concerning SNRs, there are only 3 candidates in NGC 3077
for now, but these detections were done in the radio and X ray range (Rosa-Gonzales 2005).
Chandra observations found no X ray emission originating
from prominent tidal feature (Ott et al. 2003).

\sectionb{2}{OBSERVATIONS AND DATA REDUCTION}

The observations were made on 28th February 2011 with 2 m Ritchey-Chr\`{e}tien-Coud\'{e}
telescope at the National Astronomical Observatory, Bulgaria. 
The equivalent focal length of the telescope is 16 m and the field-of-view is one
square degree with a scale 12.89"/mm. The telescope is equipped with
VersArray:1300B CCD camera with $1340$$\times1300$ px array, with a plate scale of
0".257732/px, giving a field of view $5'45"$$\times5'35"$. We used the narrow-band filters
for [SII], H$\alpha$ and red continuum (details in Table 2). We took sets of 23 images
through the [SII] and red continuum filter, and 19 images through the H$\alpha$ filter,
with 200s exposure time for each image. Typical seeing was 1.25" -– 1.75".
Standard star images, bias frames and sky flat-fields were also taken,
through each of the filters.

Standard reduction procedures including bias subtraction, trimming,
flat-field\-ing, image alignment and sky substraction were performed with the help of the IRIS
\footnote{Available from \texttt{http://www.astrosurf.com/buil/}}
software package. Afterwards, images taken through the same filters were
combined using a sigma clipping algorithm (comand \texttt{COMPOSIT}).
Since the images were taken with different total exposure
times, depending on the filters, we scaled all the images, normalizing them
to the flux of the stars in the field. The H$\alpha$ and [SII] images are then
continuum-subtracted. The continuum-subtracted H$\alpha$ image is given in Fig.2.

Sources in the continuum-subtracted H$\alpha$ image are extracted by smoothing
the image and then drawing 1$\sigma$ contours from the median value. Relative fluxes (total counts)
are then calculated using IRIS photometric tools. Finally, an astrometric reduction
of the H$\alpha$ image was performed by using U.S. Naval Observatory's USNO-A2.0
astrometric catalogue (Monet et al. 1998).


\begin{table}[!t]
\begin{center}
\vbox{\footnotesize\tabcolsep=3pt
\parbox[c]{124mm}{\baselineskip=10pt
{\smallbf\ \ Table 1.}{\small\
Data for NGC 3077 taken from SIMBAD$^\dag$.\lstrut}}
\begin{tabular}{@{\extracolsep{-0.5mm}}cccccccc@{}}
\hline
 Right ascension  & Declination & Redshift & Velocity & Distance$^\ddag$ & Angular size  & Magn. & Morph. \\
  $\alpha _{\mathrm{J2000}} $ & $\delta _{\mathrm{J2000}}$ & $z$ & $v$ [km s$^{-1}$] & $d$ [Mpc] & [ $'$ ]& &  type \\
\hline
\hline
  10 03 19.1 & +68 44 02.2  & 0.00004 & 12 & 3.83 & $2.97 \times 2.37$ &  10.9 (B) & I D \hstrut \\
\hline
\end{tabular}
}
\end{center}
\vskip-8mm

\vskip 6mm  $^\dag${\footnotesize
\texttt{http://simbad.u-strasbg.fr/simbad/}}\ \ \
$^\ddag${\footnotesize Dalcanton \& al. (2009)}

\end{table}


\begin{table}[!t]
\begin{center}
\vbox{\footnotesize\tabcolsep=3pt
\parbox[c]{124mm}{\baselineskip=10pt
{\smallbf\ \ Table 2.}{\small\
Characteristics of the narrow band filters.\lstrut}}
\begin{tabular}{c|ccc}
\hline
 Filter & $\lambda_0$ [nm] & FWHM [nm] & $\tau_{\mathrm{max}}$ [\%] \hstrut\lstrut\\
\hline
\hline
 Red cont. & 641.6 &  2.6  &  58.0  \hstrut \\
 H$\alpha$ & 657.2 &  3.2  &  86.7  \\
 $[$SII$]$ & 671.9 &  3.3  &  83.3  \\
\hline
\end{tabular}
}
\end{center}
\vskip 2mm
\end{table}


\begin{figure}[!tH]
\vbox{
\centerline{\psfig{figure=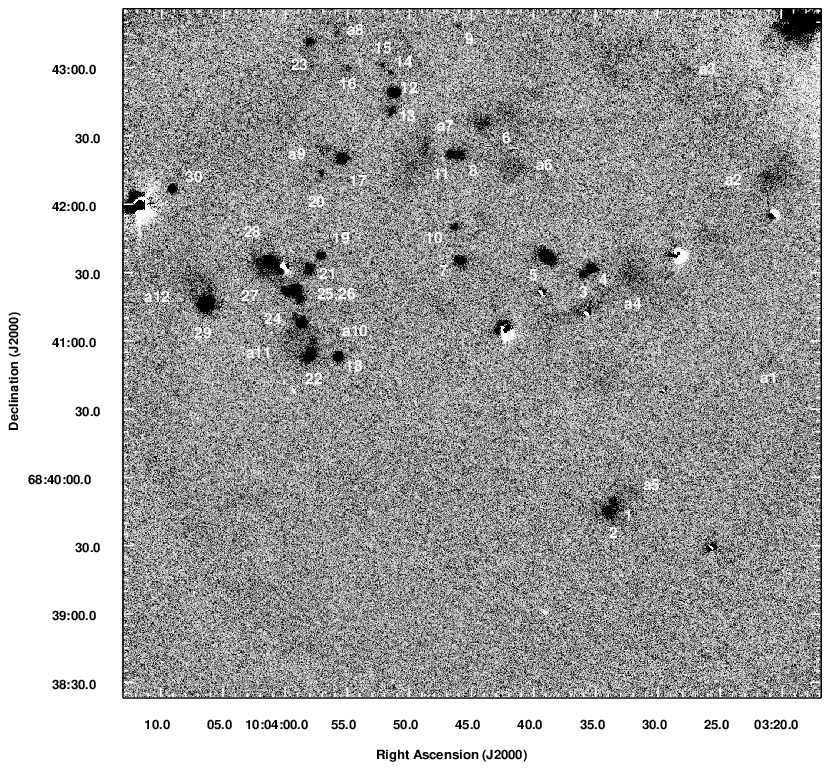,bb=178 297 414 516, width=\textwidth,angle=0,clip=}}
\vspace{1mm}
\captionb{1}
{The continuum-substracted H$\alpha$ image. New HII region candidates,
and objects used for flux calibration (Walter et al. 2006) are numbered.
Additional dark and bright features are stars not substracted well.}
}
\end{figure}


\begin{table}[!t]
\begin{center}
\vbox{\footnotesize\tabcolsep=3pt
\parbox[c]{124mm}{\baselineskip=10pt
{\smallbf\ \ Table 3.}{\small\
New HII region candidates in Garland\lstrut}}
\begin{tabular}{c|ccc}
\hline
 Source & Right ascension  & Declination &  H$\alpha$ Flux \hstrut\lstrut \\
 & $\alpha _{\mathrm{J2000}} $ & $\delta _{\mathrm{J2000}}$ &  $F_{\mathrm{H}\alpha}$ [$\times 10^{-15}$ erg cm$^{-2}$ s$^{-1}$] \\

\hline
\hline
 a1 & 10 03 20.3 & 68 40 40.1& 0.085 \hstrut \\
 a2 & 10 03 20.6 & 68 42 02.1& 2.730\\
 a3 & 10 03 27.2 & 68 42 48.7& 0.248\\
 a4 & 10 03 33.0 & 68 41 20.3& 2.129\\
 a5 & 10 03 32.4 & 68 39 44.9& 0.096\\
 a6 & 10 03 42.3 & 68 42 08.4& 0.606\\
 a7 & 10 03 49.7 & 68 42 14.1& 1.191\\
 a8 & 10 03 56.6 & 68 43 05.4& 0.332\\
 a9 & 10 03 57.6 & 68 42 13.3& 0.530\\
 a10 & 10 03 56.9& 68 40 54.9& 0.040\\
 a11 & 10 03 58.5& 68 40 50.9& 0.180\\
 a12 & 10 04 06.8& 68 41 14.6& 0.184\\
 \hline
\end{tabular}
}
\end{center}
\end{table}

\sectionb{3}{ANALYSIS AND RESULTS}

In Figure 1 we present the continuum-substracted H$\alpha$ image of the observed region.
In the H$\alpha$ filter we detected 12 new possible objects 
(marked with "a" + number in Fig. 1), with high H$\alpha$ emission presumably HII regions.
The continuum-subtracted [SII] image did not show any object with an enhanced [SII] emission,
so it appears that there were no SNR candidates.
Six newly detected sources are compact, while the other six are diffuse.
In Table 3 we give positions of these 12 sources, and their measured H$\alpha$ fluxes.
The H$\alpha$ image ($\lambda$656.3) is contaminated with some [NII] emission ($\lambda$658.3)
transmitted by the filter, so in principle the H$\alpha$ flux is actually H$\alpha$+[NII] flux.
The absolute flux calibration of the continuum-subtracted H$\alpha$ image was performed using the
fluxes of sources identified by Walter et al. (2006), as was done by Arbutina et al. (2009).

More detailed analysis about star formation rate in Garland, estimated
from H$\alpha$ emission, will be given in a future paper.

\thanks{This research has been supported by the Ministry of Education and Science of
the Republic of Serbia through project No. 176005 "Emission nebulae: structure and evolution", and project
"The Rozhen Astronomical Observatory - Major Facility for the South East European
Region" funded by UNESCO-ROSTE.}

\References

\refb Arbutina B., Ili\' c D., Stavrev K., Uro\v sevi\' c D.,
Vukoti\' c B., \& Oni\' c D. 2009, SerAJ, 179, 87

\refb Barbieri C., Bertola F., \& di Tullio G. 1974, A\&A, 35, 463

\refb Blair W.P., Long K.S. 2004, ApJS, 155, 101

\refb Dalcanton J. J. et al. 2009, ApJS, 183, 67

\refb Karachentsev I. D., Karachentseva V. E., \& Brongen F. 1985, MNRAS, 271, 731

\refb Karachentsev I. D., \& Kaisin S. S. 2007, ApJ, 133, 1883

\refb Matonick D. M., Fesen R. A. 1997, ApJS, 112, 49

\refb Monet D. et al. 1998, USNO-A2.0 - A catalog of astrometric
standards, U.S. Naval Observatory
\texttt{(http://tdc-www.harvard.edu/catalogs/ua2.html)}

\refb Ott J., Martin C. L., \& Walter F. 2003, ApJ, 594, 776

\refb Rosa-Gonzales D. 2005, MNRAS, 364, 1304

\refb Thomasson M., \& Donner K. J. 1993, A\&A, 272, 153

\refb Walter F., Martin C., \& Scoville N. 2002, AJ, 123, 225

\refb Walter F., Martin C., \& Ott J. 2006, AJ, 132, 2289

\refb Yun M.S., Ho P.T.P., \& Lo K.Y. 1994, Nature, 372, 530

\refb Yun M. S. 1999, IAUS, 186, 81

\end{document}